\documentclass[reprint,amsmath,amssymb,aps,prb,floatfix,twocolumn,superscriptaddress]{revtex4-2}
\usepackage{subfigure}
\usepackage{braket}
\usepackage{lipsum} 
\usepackage[colorlinks=true]{hyperref}
\usepackage{amsmath}
\usepackage{slashed}
\usepackage{xcolor}
\usepackage{graphicx}
\usepackage{dcolumn}
\usepackage{bm}
\usepackage{orcidlink}
\allowdisplaybreaks
\begin{document}

\title{Beyond one-loop calculation: Higher-order effects on Gross-Neveu-Yukawa tensorial criticality}

\author{SangEun Han\,\orcidlink{0000-0003-3141-1964}
}
\affiliation{Department of Physics, Simon Fraser University, Burnaby, British Columbia V5A 1S6, Canada}
\author{Igor F.~Herbut\,\orcidlink{0000-0001-5496-8330}
}
\affiliation{Department of Physics, Simon Fraser University, Burnaby, British Columbia V5A 1S6, Canada}
\affiliation{Institute for Solid State Physics, University of Tokyo, Kashiwa 277-8581, Japan}
\affiliation{Institut f\"{u}r Theoretische Physik und Astrophysik, Universit\"{a}t W\"{u}rzburg, 97074 W\"{u}rzburg, Germany}
\affiliation{Würzburg-Dresden Cluster of Excellence ct.qmat, Am Hubland, 97074 W\"{u}rzburg, Germany}

\date{\today}

\begin{abstract}
We study the Gross-Neveu-Yukawa field theory for the  SO($N$) symmetric traceless rank-two tensor order parameter coupled to Majorana fermions using the $\epsilon$-expansion around upper critical dimensions of $3+1$ to two loops. Previously we established in the one-loop calculation that the theory does not exhibit a critical fixed point for $N \geq 4$, but that nevertheless the stable fixed point inevitably emerges at a large number of fermion flavors $N_f$.  For $N_f < N_{f,c1} \approx N/2$, no critical fixed point exists; for $N_{f,c1} < N_f < N_{f,c2}$, a real critical fixed point emerges from the complex plane but fails to satisfy the additional stability conditions necessary for a continuous phase transition; and finally only for $N_f > N_{f,c2} \approx N$, the fixed point satisfies the stability conditions as well. In the present  work we compute the $O(\epsilon)$ (two-loop) corrections to the critical flavour numbers $N_{f,c1} $ and $N_{f,c2}$. Most importantly, we observe a sharp decrease in $N_{f,c2}$ from its one-loop value, which brings it closer to the point $N_f =1$ relevant to the standard Gross-Neveu model. Some three-loop results are also presented and discussed.
\end{abstract}

\maketitle

\section{Introduction}

The Gross-Neveu model in 2+1 dimensions \cite{Gross1974,Zinn-Justin,Rosenstein1989,Vasilev1993,GRACEY1994a,GRACEY1994b,Erramilli2023,ZINNJUSTIN1991} provides the simplest example of fermionic quantum criticality, with applications spanning from high-energy physics to condensed matter. Its extensions in condensed matter physics have proven essential for understanding quantum phase transitions in interacting Dirac materials. What distinguishes these transitions from conventional classical phase transitions \cite{Herbut2007} is the presence of gapless Dirac fermions at the transition  point \cite{Herbut2006,Herbut2009a,Herbut2009b,Bitan2013,*Bitan2016e,Mihaila2017,Zerf2017,Ihrig2018,VOJTA2000,Huh2008,Schwab2022,Scherer2016,Torres2018,Herbut2024book,*Igor2024}. The Gross-Neveu universality classes represent a rare instance where quantum critical phenomena beyond the standard Ginzburg-Landau paradigm can be systematically  analyzed and understood. This is particularly true for the Gross-Neveu-Yukawa (GNY) field theory, in which the interaction between the Dirac fermions and bosonic order parameters as well as the order parameter's dynamics are made explicit \cite{Herbut2024book,*Igor2024}.

One of us has with Mandal recently proposed a unified description of insulating and superconducting order parameters that open the  relativistic mass-like gap in the fermion spectrum  \cite{Igor2023}. In graphene for example, this results in the above order parameters becoming the components of a symmetric traceless tensor under the group SO(8). More generally, for $N$ two-component Dirac fermions, the general order parameter is a second-rank irreducible symmetric tensor under SO($2N$). Mean-field analysis of the Gross-Neveu model, which embodies such a symmetry, suggests the existence of a phase transition at a novel critical point where SO($2N$) symmetry is not preserved as it would normally be, but becomes spontaneously broken to SO($N$)$\times$SO($N$). This possibility has been demonstrated for the specific extension of the Gross-Neveu model to a large number of fermion flavors \cite{han2024a}.

The GNY field theory for the symmetric traceless rank-two tensor field generically features two types of quartic self-interactions \cite{Pelissetto2018,han2024b}: `double-trace' and `single-trace' quartic terms. While the double-trace quartic interaction appears in the field theories for the conventional vector-like  order parameters, the single-trace quartic interaction emerges specifically from the tensor structure of the order parameters.
The interplay of these two self-interactions during the renormalization group flow determines the ultimate nature of the phase transition in the theory.
One-loop calculation suggests that the critical point and the concomitant continuous phase transition do not exist in the Gross-Neveu model (equivalent to GNY theory with $N_{f}=1$) with $2 N\geq4$, and that the SO($2N$) to SO($N$)$\times$SO($N$) transition is weakly first order \cite{han2024a,han2024b}. The situation changes however if one introduces an independent number $N_{f} >1 $ of fermion flavors \cite{han2024b}: for a fixed $N$, increasing such an independent $N_{f}$ inevitably yields a stable fixed point of the RG flow.
The one-loop critical value of $N_{f}$ for a given $N$ has been established in the previous work, suggesting that the Gross-Neveu value of $N_f=1$ lies firmly in the regime of the fluctuation-induced first-order transition for $N=2$ and $N=4$, for example \cite{han2024b}.

In this study we investigate the higher-order corrections to our results by analyzing the two-loop RG flow equations. We find that the overall dependence of the nature of the phase transition on $N$ and $N_{f}$ identified in the previous one-loop computations \cite{han2024b} remains robust in our two-loop results, allowing us to extract the $O(\epsilon) $ corrections to the critical values of $N_{f}$. Importantly, we find that $N_{f,c1}$ is increased while $N_{f,c2}$ is  decreased by the two-loop corrections, narrowing the intermediate region where the critical fixed point exists in the unstable region of the coupling space. The decreased value of $N_{f,c2}$ indicates that fewer fermions may be required for a continuous phase transitions at a given $N$ than previously thought. For the specific case of SO(8) symmetry relevant to the spin-1/2 fermions in graphene, for example, we observe approximately 34\% reduction from the one-loop results. Furthermore, we calculate the critical exponents at the next-leading order, providing a more comprehensive understanding of these quantum phase transitions.

This paper is organized as follows. In Sec.~\ref{sec:model}, we introduce the Gross-Neveu-Yukawa model for SO($N$) rank-two symmetric traceless tensor order parameters with Majorana fermions and summarize the possible symmetry-broken ground states of the order parameters. In Sec.~\ref{sec:two-loop}, we present the two-loop order RG flow equations and discuss their consequences. We extract the two-loop corrections for the critical values of $N$ in the absence of fermions and for the critical values of $N_{f}$ with fixed $N$ in the presence of fermions. Additionally, we exhibit the three-loop contributions to the critical values of $N$ and $N_{f}$ with and without fermions in Sec.~\ref{sec:three-loop}. In Sec.~\ref{sec:honeycomb}, we discuss the specific case of SO(8)-symmetric Majorana fermions relevant to electrons on the honeycomb lattice. In the final Sec.~\ref{sec:discussion}, we give a further discussion of our work. 

\section{Model}\label{sec:model}
We introduce the Gross-Neveu-Yukawa theory for the  SO($N$)-symmetric traceless rank-two tensor order parameter coupled to the Majorana fermions with the action given by \cite{han2024b,han2025a}
\begin{widetext}
\begin{align}
\mathcal{L}={}&\int d^{D}x\;\frac{1}{2}[(\bar{\psi}(\mathbb{I}_{NN_{f}}\otimes(\gamma_{a}\partial_{a})\psi)+\bar{g}(\bar{\psi}(\mathbb{I}_{N_{f}}\otimes S\otimes\mathbb{I}_{d_{\gamma}})\psi)]\notag\\
&\quad\quad+\int d\tau d^{d}x \frac{1}{\bar{N}_{\text{Tr}}}\left[\frac{1}{2}\text{Tr}[(\partial_{\tau}S)^{2}+(\nabla S)^{2}+rS^{2}]+\frac{\bar{\lambda}_{1}}{4\bar{N}_{\text{Tr}}}(\text{Tr}[S^{2}])^{2}+\frac{\bar{\lambda}_{2}}{4}\text{Tr}[S^{4}]\right],
\end{align}
\end{widetext}
where $\psi$ is $d_{\gamma}NN_{f}$ component Majorana fermion, $\bar{\psi}\equiv\psi^{\intercal}\gamma_{0}$, $S=\sum_{i=1}^{N_{s}}\varphi_{i}\mathbb{S}^{i}$, and $\varphi_{i}$ is order parameter. Here, $S$ transforms as $S\rightarrow OSO^{\intercal}$ where $O\in \text{SO}(N)$, and $\mathbb{S}^{i}$ is $N\times N$ symmetric traceless matrix with $\text{Tr}[\mathbb{S}^{i}\mathbb{S}^{j}]=\bar{N}_{\text{Tr}}\delta_{ij}$ ($i,j=1,\cdots,N_{s}=(N-1)(N+2)/2$). 
$\gamma_{a}$ ($a=0,\cdots D-1$) satisfies the Clifford algebra, $\{\gamma_{a},\gamma_{b}\}=2\delta_{ab}\mathbb{I}_{d_{\gamma}}$ where $d_{\gamma}=2^{\lfloor D/2\rfloor}$, and we assume $\gamma_{0}$ is antisymmetic.
Note that we will set $d_{\gamma}=2$ at the end of the computation because we consider $D=2+1$ dimensions, so $\psi$ is $2NN_{f}$ component Majorana fermion.
The parameters $\bar{g}$ and $\bar{\lambda}_{1,2}$ are bare coupling constants for the Yukawa interaction and quartic interactions, respectively, which have the engineering scaling dimension $4-D$.

The model features two types of self-interactions allowed by the symmetry: single-trace and double-trace quartic interactions. 
When $N\geq4$, two quartic terms are mutually independent. The presence of two quartic interactions is responsible for a non-trivial RG flows and the possibility of the fluctuation-induced first-order transition in the theory \cite{Pelissetto2018,Janssen2022,han2024b}.
Note that the cubic term, $\text{Tr}[S^3]$, is absent, due to the additional inversion ($\mathbb{Z}_2$) symmetry that exists in the presence of fermions: $S \rightarrow -S$, $\psi \rightarrow i\gamma_0 \gamma_1 \psi$, $\partial_2 \rightarrow -\partial_2$, for example, leaves the action invariant \cite{han2024b}.

For negative $\bar{r}$, the tensor-valued order parameter can assume a variety of the ordered ground state depending on the signs of $\bar{\lambda}_{1,2}$ and the value of $N$: (a) for $\bar{\lambda}_{1}>0$ and $\bar{\lambda}_{2}<0$, the minimum of the action is at 
\begin{align}
\bar{S}=\bar{\varphi}_{0}\left(\begin{matrix}
\mathbb{I}_{N-1}&0\\
0&-(N-1)
\end{matrix}\right),
\end{align}
where $\bar{\varphi}_{0}$ is the real-valued order parameter amplitude. 
In this case, SO($N$) symmetry is broken to SO($N-1$). The stability of the theory also demands that the following condition is satisfied:
\begin{align}
\bar{\lambda}_{1}+\frac{(N^{2}-3N+3)}{(N-1)}\left(\frac{\bar{N}_{\text{Tr}}}{N}\right)\bar{\lambda}_{2}>0.
\end{align}
(b) When $\bar{\lambda}_{2}>0$, regardless of the sign of $\bar{\lambda}_{1}$, and for even $N$, the action has the following minimum
\begin{align}
\bar{S}=\bar{\varphi}_{0}\left(\begin{matrix}
\mathbb{I}_{N/2}&0\\
0&-\mathbb{I}_{N/2}\\
\end{matrix}\right)
\end{align}
and $\bar{\lambda}_{1,2}$ need to satisfy the following inequality for stability:
\begin{align}
\bar{\lambda}_{1}+\frac{\bar{N}_{\text{Tr}}}{N}\bar{\lambda}_{2}>0.\label{eq:stab_even}
\end{align}
Here, SO($N$) symmetry is broken to SO($N/2$)$\otimes$SO($N/2$).
For completeness, we also consider odd $N$, when the minimum is given by
\begin{align}
\bar{S}=\bar{\varphi}_{0}\left(\begin{matrix}
\mathbb{I}_{(N+1)/2}&0\\
0&-\frac{(N+1)}{(N-1)}\mathbb{I}_{(N-1)/2}\\
\end{matrix}\right),
\end{align}
and $\bar{\lambda}_{1,2}$ need to satisfy:
\begin{align}
\bar{\lambda}_{1}+\frac{(N^{2}+3)}{(N^{2}-1)}\left(\frac{\bar{N}_{\text{Tr}}}{N}\right)\bar{\lambda}_{2}>0.\label{eq:stab_odd}
\end{align}
At the minimum, SO($N$) is reduced to SO($(N+1)/2$)$\otimes$SO($(N-1)/2$).

It is known that the RG exhibits no critical fixed point for $N\geq4$ without fermions \cite{Pelissetto2018,han2024b}. By introducing the Yukawa interaction to a large flavor $N_{f}$ of fermions, the RG eventually shows a critical fixed point with $\bar{\lambda}_{1}>0$ and $\bar{\lambda}_{2}<0$ in the one-loop computation \cite{han2024b}.%

For completeness, we also note that for $N=2$ and $N=3$, $\text{Tr}[S^{4}]=(\text{Tr}[S^{2}])^{2} / 2$, so the two quartic interactions can both be written as the double-trace quartic interaction alone. Due to this fact the symmetric tensor order parameter can be considered as if belonging to the vector representation of SO(2) and SO(5),  respectively, and the theory has the continuous phase transition regardless of $N_{f}$. However, the critical behavior of the tensor theory when $N=3$ is distinct from that of the vector theory when both are coupled to Dirac fermions \cite{han2025a}. The details are discussed in Ref.~\cite{han2025a}. 

\section{Two-loop RG flow equations}\label{sec:two-loop}

We compute the two-loop order RG flow equations around $D=4$ space-time dimensions by using the dimensional regularization with the modified minimal subtraction ($\overline{\text{MS}}$) scheme \cite{Jack2024,han2025a}. As a check of our computation we have also used \texttt{RGBeta} \cite{Thomsen2021,SuppMat} which automates this calculation, and found the consistent results. The two-loop order RG flow-equations are given by
\begin{widetext}
\begin{align}
\frac{d\alpha_{g}}{d\ell}={}&\epsilon\alpha_{g}-\frac{(N^{2}+(N_{f}+3)N-6)}{8}\alpha_{g}^{2}+\frac{(N^{4}+2(6N_{f}+1)N^{3}+3(8N_{f}-9)N^{2}-12(4N_{f}+3)N+36)}{512}\alpha_{g}^{3} \notag\\
&+\left(\frac{(2N^{2}+3N-6)}{16}\lambda_{1}+\frac{(N^{4}+5N^{3}-6N^{2}-36N+72)}{64}\lambda_{2}\right)\alpha_{g}^{2} \notag\\
&
-\left(\frac{(N^{2}+N+2)}{64}\lambda_{1}^{2}+\frac{(2N^{2}+3N-6)}{32}\lambda_{1}\lambda_{2}+\frac{(N^{4}+5N^{3}-6N^{2}-36N+72)}{256}\lambda_{2}^{2}\right)\alpha_{g},\label{eq:RG1}\\
\frac{d\lambda_{1}}{d\ell}={}&\epsilon\lambda_{1}-\frac{NN_{f}}{4}\alpha_{g}\lambda_{1}-\frac{(N^{2}+N+14)}{8}\lambda_{1}^{2}-\frac{(2N^{2}+3N-6)}{4}\lambda_{1}\lambda_{2}-\frac{3(N^{2}+6)}{8}\lambda_{2}^{2} \notag\\
&-\frac{N^{3}N_{f}}{32}\alpha_{g}^{3}+\left(\frac{(N+2)(3N-2)}{4}\lambda_{1}-(N^{2}+6)\lambda_{2}\right)\frac{NN_{f}}{32}\alpha_{g}^{2} \notag\\
&+\left(\frac{(N^{2}+N+14)}{2}\lambda_{1}^{2}+(2N^{2}+3N-6)\lambda_{1}\lambda_{2}+\frac{3(N^{2}+6)}{2}\lambda_{2}^{2}\right)\frac{NN_{f}}{32}\alpha_{g} \notag\\
&+\frac{3(3N^{2}+3N+22)}{32}\lambda_{1}^{3}+\frac{11(2N^{2}+3N-6)}{16}\lambda_{1}^{2}\lambda_{2}+\frac{(5N^{4}+25N^{3}+114N^{2}-180N+1224)}{128}\lambda_{1}\lambda_{2}^{2} \notag\\
&+\frac{3(2N^{4}+7N^{3}-18N^{2}+36N-114)}{64}\lambda_{2}^{3},\label{eq:RG2}\\
\frac{d\lambda_{2}}{d\ell}={}&\epsilon\lambda_{2}-\frac{NN_{f}}{4}\alpha_{g}\lambda_{2}+\frac{NN_{f}}{4}\alpha_{g}^{2}-3\lambda_{1}\lambda_{2}-\frac{(2N^{2}+9N-36)}{8}\lambda_{2}^{2} \notag\\
&-\frac{NN_{f}(N^{2}+4N-8)}{32}\alpha_{g}^{3}-\left(\lambda_{1}-\frac{(3N^{2}-7N+38)}{16}\lambda_{2}\right)\frac{NN_{f}}{8}\alpha_{g}^{2}+\left(3\lambda_{1}+\frac{(2N^{2}+9N-36)}{8}\lambda_{2}\right)\frac{NN_{f}}{8}\lambda_{2}\alpha_{g} \notag\\
&+\frac{(5N^{2}+5N+154)}{32}\lambda_{1}^{2}\lambda_{2}+\frac{(22N^{2}+69N-246)}{16}\lambda_{1}\lambda_{2}^{2}+\frac{3(N^{4}+13N^{3}+2N^{2}-276N+696)}{128}\lambda_{2}^{3}.\label{eq:RG3}
\end{align}
\end{widetext}
where $\epsilon=4-D$, and $\alpha_{g}$ and $\lambda_{1,2}$ are dimensionless parameters defined as
\begin{align}
\alpha_{g}=\frac{\bar{N}_{\text{Tr}}}{N}\frac{\mu^{\epsilon}\bar{g}^{2}}{2\pi^{2}},\quad
\lambda_{1}={}\frac{\mu^{\epsilon}\bar{\lambda}_{1}}{2\pi^{2}},\quad
\lambda_{2}={}\frac{\bar{N}_{\text{Tr}}}{N}\frac{\mu^{\epsilon}\bar{\lambda}_{2}}{2\pi^{2}},
\end{align}
and $\ell=\ln(\Lambda/\mu)$. Here, we set $\bar{r}=0$ and $d_{\gamma}=2$ to get the result for $D=3$ case.
Note that by replacing $N_{f}\rightarrow 2N_{f}$ one gets the flow-equations in $D=4$ because $d_{\gamma}=2$ in $D=3$ but $d_{\gamma}=4$ in $D=4$.
Also, one can obtain the flow-equations for $N=2,3$  by simply considering the flow of  $\lambda=\lambda_{1}+(N/2)\lambda_{2}$, because, as we already mentioned, the two quartic ivariants are in this case proportional \cite{han2024b,han2025a}.

\subsection{Critical $N$ without fermions}
Let us begin by analyzing the theory in the absence of fermions first. In this case one finds two critical values of $N$, $N_{c1}$ and $N_{c2}$:
below $N_{c1}$, the Wilson-Fisher fixed point at $\lambda_2 =0$ is stable and plays a role of the critical fixed point. At $N_{c1}$, the Wilson-Fisher fixed point collides with another unstable fixed point, and they exchange stability with each other. The Wilson-Fisher fixed point is then no longer the critical fixed point, and the critical fixed point moves to the negative $\lambda_{2}$ region as $N$ increases. Above $N_{c2}$, the critical fixed point disappears due to its collision with another fixed point \cite{Pelissetto2018,han2024b}. In our previous study \cite{han2024b}, we found the one-loop results for these critical values of $N$, $N_{c1}\approx(\sqrt{41}-1)/2\approx2.702$ and $N_{c2}\approx3.624$.
From the present two-loop RG flow-equations Eqs.~\eqref{eq:RG2} and \eqref{eq:RG3}, we can now extract their next-order correction $\sim \epsilon$ to be as follows \cite{Igor1997,Gracey2018,Ihrig2019}:
\begin{align}
N_{c1}\approx{}&2.702-0.625\epsilon,\\
N_{c2}\approx{}&3.624-0.089\epsilon.
\end{align}
For example, this yields $N_{c1}\approx 2.077$ and $N_{c2}=3.535$ for $\epsilon=1$. This is consistent with a previous study on the antiferromagnetic RP model \cite{Pelissetto2018}.
Therefore, in the absence of fermions, $N_{c1}$ and $N_{c2}$ both decrease in the two-loop order computation, with $N=2$ and $N=3$ theories still possessing critical fixed points \cite{han2025a}, whereas  $N\geq4$ is not.
\begin{figure}[t]
\subfigure[]{
\includegraphics[width=0.47\linewidth]{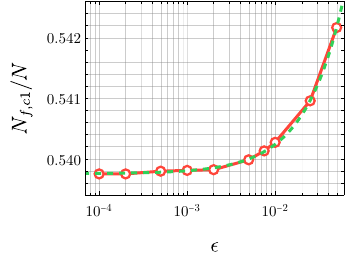}
}\hfill
\subfigure[]{
\includegraphics[width=0.47\linewidth]{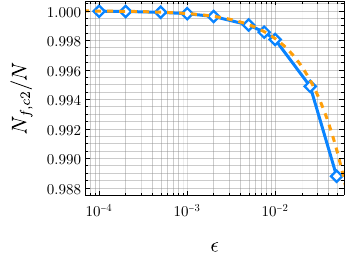}
}
\caption{The plots for $N_{f,c1}/N$ and $N_{f,c2}/N$ as a function of $\epsilon$ in the $N\rightarrow\infty$ limit. (a) The plot for $N_{f,c1}/N$ and (b) the plot for $N_{f,c2}/N$. The red circles and blue diamonds stand for the $N_{f,c1}/N$ and $N_{f,c2}/N$ from the RG flow equations for given $\epsilon$ and $N\rightarrow\infty$ limit. The dashed lines are the fitting values from Eqs.~\eqref{eq:nfc1} and \eqref{eq:nfc2}.}\label{fig:2loop_Nf_plot}
\end{figure}

\begin{figure*}
\subfigure[]{
\includegraphics[width=0.23\linewidth]{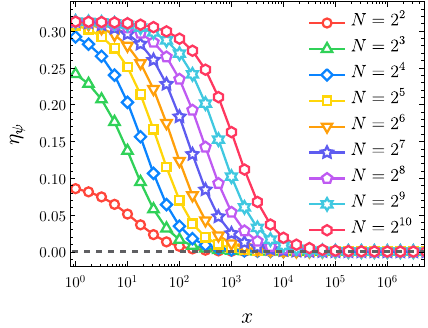}
}
\subfigure[]{
\includegraphics[width=0.23\linewidth]{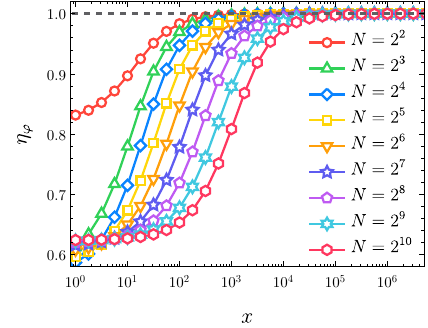}
}
\subfigure[]{
\includegraphics[width=0.23\linewidth]{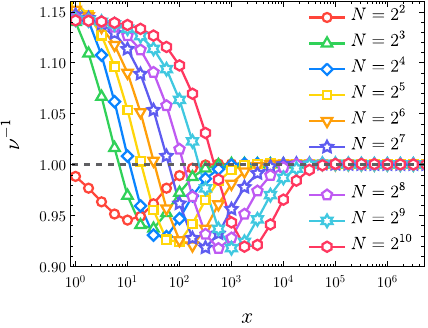}
}
\subfigure[]{
\includegraphics[width=0.23\linewidth]{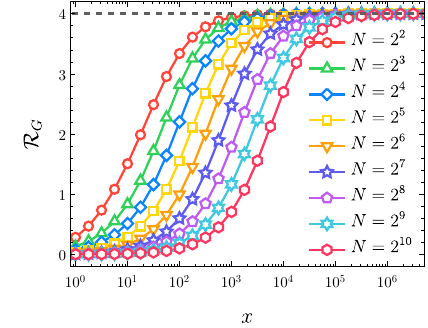}
\label{fig:crit_exp_d}
}
\caption{The two-loop critical exponents and mass gap ratio in two-loop order with $\epsilon=1$. (a-b) The anomalous dimensions of fermions and order parameters, $\eta_{\psi}$ and $\eta_{\varphi}$, (c) the inverse correlation exponents $\nu^{-1}$, and the mass gap ratio.}\label{fig:crit_exp}
\end{figure*}

\subsection{Critical $N_{f}$ with fermions}
Let us consider the case with the fermions included next.
In the presence of fermions, for fixed $N\geq4$, we distinguish three regimes depending on $N_{f}$, and find two critical values of $N_{f}$, $N_{f,c1}$ and $N_{f,c2}$.  First, there is no stable fixed point when $N_{f}<N_{f,c1}$. 
For $N_{f}\geq N_{f,c1}$, the fixed point with $\alpha_{g}^{*}>0$, $\lambda_{1}^{*}<0$, and $\lambda_{2}^{*}>0$ emerges, but it fails to  satisfy the condition for the continuous phase transition mentioned above (Eqs.~\eqref{eq:stab_even} and \eqref{eq:stab_odd}. In this regime we therefore still expect a  fluctuation-induced first order transition.
Increasing $N_{f}$ drives the fixed point toward the vertical axis. Finally, for $N_{f,c1}<N_{f,c2}\leq N_{f}$, the RG flow equations have a stable fixed point which also satisfies the stability condition.  This behavior persists both in the one-loop and the two-loop calculation.
To quantify these observations we examine the critical values of $N_{f}$, $N_{f,c1}$ and $N_{f,c2}$.

At one-loop, $N_{f,c1}$ and $N_{f,c2}$ are given by $N_{f,c1}\approx(C_{c1}/2)N$ and $N_{f,c2}\approx N-2$ where $C_{c1}\approx1.080$ \cite{han2024b}.
At two-loops, the RG flow equations Eqs.~\eqref{eq:RG1}-\eqref{eq:RG3}, in the large $N$ limit, yield the leading $\epsilon$ corrections as follows \cite{Igor1997,Gracey2018,Ihrig2019}:
\begin{align}
N_{f,c1}\approx{}&(0.540+0.048\epsilon)N,\label{eq:nfc1}\\
N_{f,c2}\approx{}&(N-2)-(0.188N-0.435)\epsilon+\mathcal{O}(\epsilon^{2}).\label{eq:nfc2}
\end{align}
We present $N_{f,c1}/N$ and $N_{f,c2}/N$ in terms of $\epsilon$ in $N\rightarrow\infty$ limit in Fig.~\ref{fig:2loop_Nf_plot}. Evidently, the two-loop order contribution reduces the value of $N_{f,c2}$ for given $N$ and $\epsilon$.
Note that for the odd $N$ case, there is an additional prefactor $(N^{2}+3)/(N^{2}-1)$ (see Eq.~\ref{eq:stab_odd}), but this approaches 1 in the large $N$ limit. Therefore, we can use the same $N_{f,c2}$ for both even and odd values of $N$ in this limit.

By taking naive extrapolation and putting $\epsilon=1$, for large $N$, we have
\begin{align}
N_{f,c1}\approx{}&0.588N,\\
N_{f,c2}\approx{}&0.812N,
\end{align}
We notice that  $N_{f,c1}$ is higher, whereas $N_{f,c2}$ is lower than the one-loop results.
Furthermore, since they are still both linear in $N$, they can never cross, and the existence of the three regimes persists in the large $N$ limit.

\subsection{Universal quantities to two-loops}

We compute the two-loop values of the critical exponents next. The expressions in terms of $\alpha_{g}$ and $\lambda_{1,2}$ for the anomalous dimensions of fermions and bosons, $\gamma_{\psi}$ and $\gamma_{\phi}$ , and the inverse correlation length exponent $\nu^{-1}$ are given by
\begin{widetext}
\begin{align}
\eta_{\psi}={}&\frac{(n-1)(n+2)}{16}\alpha_{g}-\frac{(n-1)(n+2)(N^{2}+(6N_{f}+1)N-2)}{1024}\alpha_{g}^{2},\label{eq:critexp1}\\
\eta_{\phi}={}&\frac{NN_{f}}{8}\alpha_{g}-\frac{(3N^{2}+5N-10)NN_{f}}{256}\alpha_{g}^{2}
+\frac{(N^{2}+N+2)}{64}\lambda_{1}^{2}
+\frac{(2N^{2}+3N-6)}{32}\lambda_{1}\lambda_{2}\notag\\&+\frac{(N^{2}+5N^{3}-6N^{2}-36N+72)}{256}\lambda_{2}^{2},\label{eq:critexp2}\\
\nu^{-1}={}&
2-\frac{NN_{f}}{8}\alpha_{g}-\frac{(N^{2}+N+2)}{8}\lambda_{1}-\frac{(2N^{2}+3N-6)}{8}\lambda_{2}+\frac{NN_{f}(N+1)(3N-2)}{256}\alpha_{g}^{2}
\notag\\&+\left(\frac{(N^{2}+N+2)}{8}\lambda_{1}+\frac{(2N^{2}+3N-6)}{8}\lambda_{2}\right)\frac{NN_{f}}{8}\alpha_{g}
+\frac{5(N^{2}+N+2)}{64}\lambda_{1}^{2}+\frac{5(2N^{2}+3N-6)}{32}\lambda_{1}\lambda_{2}\notag\\&+\frac{5(N^{4}+5N^{3}-6N^{2}-36N+72)}{256}\lambda_{2}^{2}.\label{eq:critexp3}
\end{align}
\end{widetext}
By substituting the fixed-point values into the above expressions the corresponding actual critical exponents can be obtained.
For given $N$, the critical exponents in terms of $x=N_{f}-N_{f,c2}$ with $\epsilon=1$ are presented in Fig.~\ref{fig:crit_exp}.
Furthermore, we can consider the mass gap ratio, 
\begin{align}
\mathcal{R}_{G}=(\bar{N}_{\text{Tr}}/N)\frac{m_{\varphi}^{2}}{m_{\psi}^{2}}=\frac{2(\lambda_{1}+\lambda_{2})}{\alpha_{g}},
\end{align}
for even $N$, and
\begin{align}
\mathcal{R}_{G}=(\bar{N}_{\text{Tr}}/N)\frac{m_{\varphi}^{2}}{m_{\psi}^{2}}=\frac{2(\lambda_{1}+\frac{(N^{2}+3)}{(N^{2}-1)}\lambda_{2})}{\alpha_{g}},
\end{align}
for odd $N$. In the odd $N$ case, there are two $m_{\psi}^{2}$ values and we choose larger one for the mass gap ratio. 
Substituting the fixed-point value into the above expression yields the results for the mass gap ratio. We present the mass gap ratio with even $N$ in Fig.~\ref{fig:crit_exp_d}.

As an illustrative example, we fit the quantities for $N=8$ to the following two-loop expressions in $x$,
\begin{widetext}
\begin{align}
\eta_{\psi}(N=8,x)\approx{}&\frac{4.375}{x+21.048}\epsilon
-\frac{3.276 x-28.601}{x^2+44.609 x+543.391}\epsilon^{2}+\mathcal{O}(\epsilon^{3}),\label{eq:quanti1}\\
\eta_{\varphi}(N=8,x)\approx{}&\frac{x+10.798}{x+21.048}\epsilon
+\frac{7.423 x+33.394}{x^2+39.042 x+492.367}\epsilon^{2}+\mathcal{O}(\epsilon^{3}),\label{eq:quanti2}\\
\nu^{-1}(N=8,x)\approx{}&2-
\frac{x^2+110.740 x+730.142}{x^2+85.332 x+744.264}\epsilon+
\frac{28.110 x^2+131.226 x+3981.517}{x^3+154.332 x^2+3328.534 x+23549.943}\epsilon^{2}+\mathcal{O}(\epsilon^{3}),\label{eq:quanti3}\\
\mathcal{R}_{G}(N=8,x)\approx{}&\frac{4x^3+414.006 x^2+5394.499 x}{x^3+186.546 x^2+7129.540 x+42605.446}\notag\\
&+\frac{214.652 x^3+450.629 x^2+7451.347 x}{x^4+269.824 x^3+14536.746 x^2+151977.558 x+455734.668}\epsilon+\mathcal{O}(\epsilon^{2}).\label{eq:quanti4}
\end{align}
Note that here $\mathcal{R}_{G}$ is given up to $\epsilon$-linear corrections because the one-loop computation gives us $\epsilon^{0}$-order values \cite{ZINNJUSTIN1991,han2018,han2024b}.
We also present, in Fig.~\ref{fig:majo_plot}, a comparison among the one-loop and two-loop results and their fitted values shown in Eqs.~\eqref{eq:quanti1}--\eqref{eq:quanti4} for each quantity at $N=8$ with $\epsilon=1$.
Note that, as exemplified in the case of $N=8$, the functional forms used in Eqs.~\eqref{eq:quanti1}--\eqref{eq:quanti4} provide good fits for various values of $N$ and $x$ as well.

For the limits of $x=N_{f}-N_{f,c2}(N)\gg N\gg1$,
\begin{align}
    \eta_{\psi}(N,x)\approx{}&\frac{N/2}{x+2N}\epsilon-\frac{0.374 N x -0.308 N^2}{x^2 + 4.209 N x + 4.926 N^2}\epsilon^{2}+\mathcal{O}(\epsilon^{3}),\\
    \eta_{\varphi}(N,x)\approx{}&\frac{x+N}{x+2N}\epsilon+\frac{0.764 N x + 0.555 N^2}{x^2 + 3.656 N x + 4.443 N^2}\epsilon^{2}+\mathcal{O}(\epsilon^{3}),\\
    \nu^{-1}(N,x)\approx{}&2-\frac{x^2 + 11.241 N x + 7.649 N^2}{x^2 + 8.368 N x + 7.649 N^2}\epsilon+\frac{3.203 N x^2 + 1.779 N^2 x + 3.073 N^3}{x^3 + 15.686 N x^2 + 34.351 N^2 x + 21.856 N^3}\epsilon^{2}+\mathcal{O}(\epsilon^{3}),\\
    \mathcal{R}_{G}(N,x)\approx{}&\frac{4 x^3 + 39.163 N x^2 + 50.096 N^2 x}{x^3 + 18.772 N x^2 + 75.063 N^2 x + 50.318 N^3}\notag\\&+\frac{24.021 N x^3 + 5.353 N^2 x^2 + 7.339 N^3 x}{x^4 + 27.826 N x^3 + 171.872 N^2 x^2 + 203.723 N^3 x + 79.484 N^4}\epsilon+\mathcal{O}(\epsilon^{2}).
\end{align}

\begin{figure*}
\subfigure[]{
\includegraphics[width=0.23\linewidth]{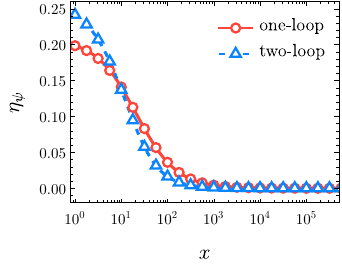}
}
\subfigure[]{
\includegraphics[width=0.23\linewidth]{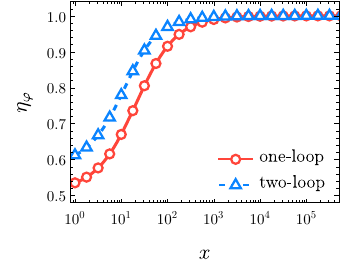}
}
\subfigure[]{
\includegraphics[width=0.23\linewidth]{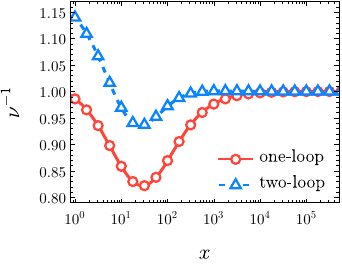}
}
\subfigure[]{
\includegraphics[width=0.23\linewidth]{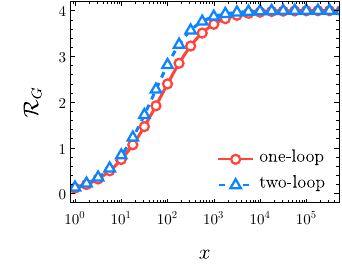}
}
\caption{The comparison between one-loop and two-loop critical exponents and mass gap ratio in terms of $x=N_{f}-N_{f,c2}$, and fitted values from Eqs.~\eqref{eq:quanti1}-\eqref{eq:quanti4} when $N=8$ with $\epsilon=1$.
The red circle and blue triangle stand for the one-loop and two-loop results, respectively.
And the red solid line and blue dashed line stand for the one-loop and two-loop fitted values from Eqs.~\eqref{eq:quanti1}-\eqref{eq:quanti4}, respectively.
}\label{fig:majo_plot}
\end{figure*}
\end{widetext}

\subsection{Honeycomb lattice with SO(8)-symmetric Majorana fermions}\label{sec:honeycomb}

One realization of the  GNY field theory under consideration arises at the half-filled honeycomb lattice of graphene \cite{Igor2023}. By representing Dirac fermions in terms of Majoranas and by utilizing Fierz identities, one can construct the GNY field theory for the SO(8) symmetric rank-2 tensor-valued order parameters \cite{Igor2023}.
Since Eq.~\eqref{eq:nfc2} is valid only in the large $N$ limit, its applicability to small values of $N$ is limited. 
Without taking large-$N$ limit, the one-loop result for graphene, where $N=8$, gives $N_{f,c2} \approx 10.798$ \cite{han2024b}. From the two-loop order computation, when $N=8$, $N_{f,c2}$ is given by
\begin{align}
N_{f,c2}(N=8)\approx{}&10.798-3.619\epsilon+\mathcal{O}(\epsilon^{2}).
\end{align}
Therefore, for $\epsilon=1$ we find $N_{f,c2}\approx7.178$, that is about a 34\% reduction from the one-loop result. 

\section{Three-loop corrections}\label{sec:three-loop}

One can also compute the three-loop contribution to the RG flow-equations and universal quantities (see Appendix~\ref{app:3loop_RG} and \ref{app:3loop_exp}). These three-loop corrections contribute to further $\epsilon^{2}$-terms in the critical values of $N$ and $N_{f}$, without and with fermions, respectively.

First, without fermions, we find the following next-order corrections to $N_{c1,c2}$,
\begin{align}
N_{c1}\approx{}&2.702- 0.625\epsilon + 0.790 \epsilon^{2}+\mathcal{O}(\epsilon^{3}),\\
N_{c2}\approx{}&3.624-0.089\epsilon+0.250\epsilon^{2}+\mathcal{O}(\epsilon^{3}).
\end{align}
Again, these are consistent with the previous study \cite{Pelissetto2018}. 
The series as usual does not appear to be  convergent, and one needs a resummation \cite{Pelissetto2018}. From the Borel-Pad\'{e} approximation, which is a combination of the Borel summation and Pad\'{e} approximation \cite{LeGuillou1977,goettkerschnetmann1999,Zinn-Justin} (for details, see Appendix~\ref{app:borel-pade}), for example, for $\epsilon=1$, we can obtain $N_{c1}=2.387$ and $N_{c2}=3.595$.

With fermions, and in the large $N$ limit, $N_{f,c2}$ up to order $\epsilon^{2}$ is given by
\begin{align}
N_{f,c2}\approx{}&(N-2)-(0.188N-0.435)\epsilon \notag\\
&+(0.719N-2.797)\epsilon^{2}+\mathcal{O}(\epsilon^{3}).
\end{align}
Evidently, the coefficient of $\epsilon^{2}$-term is considerably larger than that of $\epsilon$-term. This indicates that the $\epsilon$-expansion fails and we need a proper resummation method, just  as in the case without fermions. By using the Borel-Pad\'{e} approximation, we obtain
\begin{align}
N_{f,c2}(\epsilon=1,N)\approx0.948N-1.931+\mathcal{O}(N^{-1}).
\end{align}
While still lower than the one-loop result, $N_{f,c2}\approx N$, we find it to be considerably less so than in the two-loop calculation.

In the case of the graphene, where $N=8$, without taking the large-$N$ limit, we obtain
\begin{align}
N_{f,c2}(N=8)\approx{}&10.798 - 3.619 \epsilon + 4.640 \epsilon^{2}+\mathcal{O}(\epsilon^{3}).
\end{align}
Just as before, the $\epsilon$-expansion appears to break down. Using the Borel-Pad\'{e} approximation, $N_{f,c2}(\epsilon=1,N=8)\approx8.988$. 
This number is higher than in the two-loop calculation, and represents only about 17\% reduction from the leading (one-loop) result.

Note that despite the apparent non-convergence of the computed low-order terms, the alternating signs indicate possible eventual convergence to a critical value smaller than the one-loop estimate.

We also present the three-loop corrections in terms of $N$ and $x=N_{f}-N_{f,c2}(N)$ for the universal quantities in Appendix~\ref{app:3loop_exp}

\section{Summary and Discussion}\label{sec:discussion}

In this work we have investigated the GNY theory for SO($N$) symmetric traceless matrix-valued order parameters with $N_{f}$ species of Majorana fermions in 2+1 dimensions by computing two-loop  RG flow equations. Our analysis reveals that the critical values of $N$ without fermions decrease, yet systems with $N\geq4$ still lack a stable RG fixed point, in agreement with the one-loop analysis. In the presence of fermions, we confirm that the structure of the RG flow characterized by two critical values of the fermion flavors $N_{f,c1}$ and $N_{f,c2}$ persists beyond the leading-order computation. Importantly, for a given $N$, we discover that $N_{f,c1}$ increases while $N_{f,c2}$ decreases. We also present the critical exponents and mass gap ratio in $\mathcal{O}(\epsilon^{2})$ derived from our two-loop computation.

For the specific case of SO(8)-symmetric Majorana fermions on the honeycomb lattice, we found that $N_{f,c2}$ decreases by approximately 34\% under the two-loop calculation. The predictions presented here, including $N_{f,c2}$, critical exponents, and mass gap ratio, offer testable outcomes for future studies employing various methodologies such as numerical simulations \cite{Liu2024}, conformal bootstrapping \cite{Reehorst2023}, and experimental investigations \cite{Liguo2024,Dongyang2025}. Furthermore, our work suggests a systematic reduction of $N_{f,c2}$ in higher-order calculations so that continuous phase transitions in the systems with tensor order parameter may be more accessible than previously thought. 

To further validate our findings, we have extended our analysis to include three-loop results for $N_{f,c2}$. Notably, the coefficient of $\epsilon^{2}$ exceeds that of $\epsilon$, highlighting a common feature of the $\epsilon$-expansion. By employing Borel-Pad\'{e} approximation, we confirm that the three-loop order $N_{f,c2}$ remains lower than the one-loop result, reinforcing the trend observed in our two-loop calculations. However, comprehensive validation would ultimately require comparison with numerical simulations and other methods.

Although we have only a few low-order terms with coefficients that suggest a non-convergent series, necessitating resummation methods, the alternating signs of these coefficients suggest that the epsilon expansion may be asymptotic. Since the two-loop correction to $N_{f, c2}$ is  negative we expect that its ultimate value is still smaller than the leading-order result.

Finally, while our $\epsilon$-expansion results require resummation techniques, alternative approaches such as the Monte-Carlo study or the conformal bootstrap may provide complementary perspectives and lead to better quantitative accuracy of critical flavor numbers and critical exponents in the future.

\begin{acknowledgments}
This work was supported by the NSERC of Canada.
\end{acknowledgments}

\appendix
\section{Three loop contributions in RG flow equations}\label{app:3loop_RG}
In this section, we will present the three-loop contributions in the RG flow equations. By adding the following corrections into the two-loop RG flow equations Eqs.~\eqref{eq:RG1}-\eqref{eq:RG3} in the main text, we obtain the three-loop order RG flow equations. The three-loop corrections in the RG flow equations are given by
\begin{widetext}
\begin{align}
\Delta_{\alpha_{g}}^{(3)}={}&
-\frac{N_{f}^{2}N^{2}(13N^{2}+14N-28)}{8192}\alpha_{g}^{4}\notag\\
&-\frac{N_{f}N(79 N^4+10(24 \zeta_{3}+17) N^3 -3 (80 \zeta_{3}+91) N^2-4(624 \zeta_{3}-61) N +20 (240 \zeta_{3}-89))}{32768}\alpha_{g}^{4}\notag\\
&+(3 N^6+ (79-48 \zeta_{3})N^5- (1440 \zeta_{3}-1223)N^4- (720 \zeta_{3}-805)N^3+2 (5232 \zeta_{3}-3169) N^2\notag\\&\quad\quad-12 (912 \zeta_{3}-697) N+8 (912 \zeta_{3}-697))\frac{1}{32768}\alpha_{g}^{4}   \notag\\
&-\left[(2N^{2}+3N-6)\lambda_{1}+\frac{(N^4+5 N^3-6 N^2-36 N+72)}{4}\lambda_{2}\right]\frac{15N_{f}N}{1024}\alpha_{g}^{3}
\notag\\
&-\left[(4 N^4+19 N^3-9 N^2-84 N+84)\lambda_{1}+\frac{(N^6+12 N^5+31 N^4-124 N^3-144
   N^2+1008 N-1008)}{4}\lambda_{2}\right]\frac{3}{512}\alpha_{g}^{3} \notag\\
& +\left[\frac{(N^{2}+N+2)}{2}\lambda_{1}^{2}+(2N^{2}+3N-6)\lambda_{1}\lambda_{2}+\frac{(N^4+5 N^3-6 N^2-36 N+72)}{8}\lambda_{2}^{2}\right]\frac{15N_{f}N}{2048}\alpha_{g}^{2} \notag\\
& -\left[\frac{(5N^4+16 N^3+319 N^2+480 N-1004)}{2}\lambda_{1}^{2}+(50 N^4+237 N^3-281 N^2-1572
   N+3012)\lambda_{1}\lambda_{2}\right.\notag\\&\quad\quad\quad\left.+\frac{(5 N^6+116 N^5+323 N^4-2356
   N^3+576 N^2+24624 N-36144)}{8}\lambda_{2}^{2}\right]\frac{1}{2048}\alpha_{g}^{2}
\notag\\
&+\left[(N^4+2 N^3+17 N^2+16 N+28)\lambda_{1}^{3}
+3(2 N^4+5 N^3+25 N^2+36 N-84)\lambda_{1}^{2}\lambda_{2}\right.\notag\\&\quad\quad\quad\left.
+3(7 N^4+27 N^3-33 N^2-144 N+252)\lambda_{1}\lambda_{2}^{2}\right.\notag\\&\quad\quad\quad\left.
+\frac{(2 N^6+19 N^5+21 N^4-270
   N^3+108 N^2+2160 N-3024)}{4}\lambda_{2}^{3}\right]\frac{1}{2048}\alpha_{g},\\
\Delta_{\lambda_{1}}^{(3)}={}&
\frac{N_{f}^{2}N^{2}(9N^{2}+44)}{1024}\alpha_{g}^{4}
+\frac{N_{f}N^{3}(5 N^2+(6 \zeta_{3}+17)N-(96\zeta_{3}+34))}{1024}\alpha_{g}^{4}  \notag\\
&-\left[\frac{(64 N^2+89 N-178)}{64}\lambda_{1}+(N^{2}+6)\lambda_{2}\right]\frac{N_{f}^{2}N^{2}}{64}\alpha_{g}^{3} \notag\\
&+\frac{N_{f}N(187 N^4+(934-240 \zeta_{3})N^3 +(1439-144 \zeta_{3})N^2 -12(377-133\zeta_{3}) N+12(377-133\zeta_{3}))}{16384}\lambda_{1}\alpha_{g}^{3}  \notag\\
   &+\frac{N_{f}N((12 \zeta_{3}+11)N^4 +(36
   \zeta_{3}+39)N^3 -24 (8
   \zeta_{3}+1) N^2+72 (3-\zeta_{3}) N-144 (3-\zeta_{3}))}{512}\lambda_{2}\alpha_{g}^{3} \notag\\
&+\left[\frac{(N^2+N+14)}{2}\lambda_{1}^{2}+(2 N^2+3 N-6)\lambda_{1}\lambda_{2}+\frac{3(N^2+6)}{2}\lambda_{2}^{2}\right]\frac{3N_{f}^{2}N^{2}}{1024}\alpha_{g}^{2} \notag\\
&-\frac{N_{f}N(41 N^{4}+ (24 \zeta_{3}+84)N^{3}+ (744 \zeta_{3}+947)N^{2}+(2016 \zeta_{3}+564)N -4 (1032 \zeta_{3}+325))}{8192}\lambda_{1}^{2}\alpha_{g}^{2}\notag\\
&-\frac{N_{f}N( (48 \zeta_{3}+76)N^4+ (288
   \zeta_{3}+179)N^3+ (552
   \zeta_{3}+35)N^2-12 (168
   \zeta_{3}+25) N+(8928 \zeta_{3}+1812))}{4096}\lambda_{1}\lambda_{2}\alpha_{g}^{2}\notag\\
&-\frac{N_{f}N((192 \zeta_{3}+187)N^4 + (936 \zeta_{3}+225)N^3-720 (4\zeta_{3}-1) N^2+90 (48 \zeta_{3}+23) N-36 (456 \zeta_{3}+187))}{8192}\lambda_{2}^{2}\alpha_{g}^{2}\notag\\
&-\left[\frac{(65N^{2}+65N+514)}{2}\lambda_{1}^{3}+81(2N^{2}+3N-6)\lambda_{1}^{2}\lambda_{2}\right]\frac{N_{f}N}{1024}\alpha_{g}\notag\\
&-\left[\frac{(11N^{4}+55N^{3}+318N^{2}-396N+3096))}{16}\lambda_{1}\lambda_{2}^{2}+(2N^{4}+7N^{3}-18N^{2}+36N-144)\lambda_{2}^{3}\right]\frac{3N_{f}N}{512}\alpha_{g}\notag\\
&-\frac{(33 N^4+66 N^3+(960 \zeta_{3}+1745)N^2 +(960 \zeta_{3}+1712) N +(6528 \zeta_{3}+8284))}{2048}\lambda_{1}^{4}\notag\\
&-\frac{(2N^{2}+3N-6)(79 N^2+79 N+6(413+256 \zeta_{3}))}{1024}\lambda_{1}^{3}\lambda_{2}\notag\\
&-[3 N^6+18 N^5+(1152 \zeta_{3}+6659)N^4 +720(8 \zeta_{3}+33) N^3 +24 (864 \zeta_{3}+233)N^2  \notag\\&\quad\quad\quad-648 (64 \zeta_{3}+183) N+7776 (32 \zeta_{3}+51)]\frac{1}{8192}\lambda_{1}^{2}\lambda_{2}^{2}\notag\\
&-[10 N^6+92 N^5+(384 \zeta_{3}+745)N^4 +(1344 \zeta_{3}+535)N^3 -6 (576 \zeta_{3}+355) N^2 \notag\\&\quad\quad\quad+36(192 \zeta_{3}+559) N -72 (384 \zeta_{3}+683))]\frac{3}{2048}\lambda_{1}\lambda_{2}^{3}\notag\\
&-[(96 \zeta_{3}+193)N^6 +3(336\zeta_{3}+551) N^5 -36 (58\zeta_{3}+21) N^4-162 (176\zeta_{3}+119) N^3 \notag\\&\quad\quad\quad+108  (984\zeta_{3}+583)N^2-136728 N+3888 (32\zeta_{3}+89)]\frac{1}{8192}\lambda_{2}^{4},\\
\Delta_{\lambda_{2}}^{(3)}={}&\frac{N_{f}^{2}N^{2}(41N^{2}+157N-490)}{4096}\alpha_{g}^{4}\notag\\
&-\frac{N_{f}N(13 N^4-(144\zeta_{3}-98)N^3 -(1152\zeta_{3}-129) N^2+(1536\zeta_{3}-20)N-(1536\zeta_{3}-20))}{8192}\alpha_{g}^{4}\notag\\
&-\frac{N_{f}^{2}N^{2}}{16}\left[\lambda_{1}+\frac{(64 N^2+217 N-818)}{256}\lambda_{2}\right]\alpha_{g}^{3}\notag\\
&+\frac{N_{f}N(N^2 (12\zeta_{3}+19)+3(12 \zeta_{3}+17)N -6 (12\zeta_{3}+17))}{128}\lambda_{1}\alpha_{g}^{3}\notag\\
&+\frac{N_{f}N(315 N^4+2(264\zeta_{3}+1139)N^3 +
   (1392\zeta_{3}-161)N^2-12(944
  \zeta_{3}+2201) N +12(2288\zeta_{3}+3257))}{16384}\lambda_{2}\alpha_{g}^{3}\notag\\
&+\frac{3N_{f}^{2}N^{2}}{256}\left[3\lambda_{1}\lambda_{2}+\frac{(2N^{2}+9N-36)}{8}\lambda_{2}^{2}\right]\alpha_{g}^{2}\notag\\
&-\frac{N_{f}N(2 N^2+2 N+(27\zeta_{3}+13))}{64}\lambda_{1}^{2}\alpha_{g}^{2}\notag\\
&-\frac{N_{f}N((96\zeta_{3}+367)N^2 +99(8 \zeta_{3}+5) N -18 (160\zeta_{3}+79))}{1024}\lambda_{1}\lambda_{2}\alpha_{g}^{2}\notag\\
&-\frac{N_{f}N(166 N^4+(432\zeta_{3}+1003)N^3 +N^2
   (1224\zeta_{3}-2165)-18 N (864
  \zeta_{3}+433)+2376 (16 \zeta_{3}+7))}{8192}\lambda_{2}^{2}\alpha_{g}^{2}\notag\\
&-\frac{N_{f}N}{1024}\left[\frac{3(11N^{2}+11N+406)}{2}\lambda_{1}^{2}\lambda_{2}+9(18N^{2}+59N-214)\lambda_{1}\lambda_{2}^{2}
+\frac{(17 N^4+277 N^3+90 N^2-6372 N+16200)}{8}\lambda_{2}^{3}\right]\alpha_{g}\notag\\
&+\frac{(13 N^4+26 N^3-(384 \zeta_{3}+775)N^2 -4(96\zeta_{3}+197) N -4
   (2496\zeta_{3}+2903))}{1024}\lambda_{1}^{3}\lambda_{1}\notag\\
&+\frac{(70 N^4-83 N^3-(9216 \zeta_{3}+11941)N^2 -18(1536 \zeta_{3}+1873) N +288 (336\zeta_{3}+403))}{2048}\lambda_{1}^{2}\lambda_{2}^{2}\notag\\
&-\frac{(2 (48\zeta_{3}+91)N^4 +16(66
  \zeta_{3}+109) N^3 -1689 N^2-27(768
  \zeta_{3}+1007) N +54 (960 \zeta_{3}+1153))}{512}\lambda_{1}\lambda_{2}^{3}\notag\\
&-[26 N^6+(216\zeta_{3}+695)N^5 +
   3(1208\zeta_{3}+1363)N^4-16866 N^3-36
   (2088\zeta_{3}+1487)N^2 \notag\\&\quad\quad+216
   (1440\zeta_{3}+2071) N-1728 (384
  \zeta_{3}+451)]\frac{1}{8192}\lambda_{2}^{4},
\end{align}
where $\zeta_{3}$ is the value of the Riemann zeta function at 3, $\zeta_{3}=\zeta(3)\equiv\sum_{n=1}^{\infty}n^{-3}\approx1.202$.

\section{Three loop contributions in critical exponenets}\label{app:3loop_exp}
In this section, we will present the three-loop contributions in the critical exponent expressions. By adding the following corrections into the two-loop order critical exponent expressions in Eqs.~\eqref{eq:critexp1}-\eqref{eq:critexp3} in the main text, we obtain the three-loop order critical exponent expressions in terms of $\alpha_{g}$ and $\lambda_{1,2}$. 
The expressions for the additional contributions from the three-loop order in the critical exponents are given by
\begin{align}
\Delta_{\eta_{\psi}}^{(3)}={}
&-\frac{N_{f}^{2}N^{2}(N-1)(N+2)}{16384}\alpha_{g}^{3}+\frac{N_{f}N(N-1)(N+2)(37N^{2}+47N-94)}{32768}\alpha_{g}^{3}\notag\\
&-\frac{(N-1)(N+2)(3 N^4-(48 \zeta_{3}-34) N^3-(144 \zeta_{3}-75) N^2+(192 \zeta_{3}-60) N-(192 \zeta_{3}-6))}{65536}\alpha_{g}^{3}\\
&+\left[(2N^{2}+3N-6)\lambda_{1}+\frac{(N^{4}+5N^{3}-6N^{2}-36N+72)}{4}\lambda_{2}\right]\frac{(N-1)(N+2)}{512}\alpha_{g}^{2}\notag\\
&-\left[\frac{(N^{2}+N+2)}{2}\lambda_{1}^{2}+(2N^{3}+3N-6)\lambda_{1}\lambda_{2}+\frac{(N^{4}+5N^{3}-6N^{2}-36N+72)}{8}\lambda_{2}^{2}\right]\frac{11(N-1)(N+2)}{4096}\alpha_{g}\notag,\\
\Delta_{\eta_{\varphi}}^{(3)}={}
&\frac{N_{f}^{2}N^{2}(16N^{2}+25N-50)}{8192}\alpha_{g}^{3}\notag\\
&+\frac{N_{f}N(5 N^4+(48 \zeta_{3}-6) N^3+(144 \zeta_{3}-95) N^2-12 (16 \zeta_{3}+7) N+(192 \zeta_{3}+84) )}{32768}\alpha_{g}^{3}\notag\\
&+\left[(2N^{2}+3N-6)\lambda_{1}+\frac{(N^{4}+5N^{3}-6N^{2}-36N+72)}{4}\lambda_{2}\right]\frac{5N_{f}N}{1024}\alpha_{g}^{2}\notag\\
&-\left[\frac{(N^{2}+N+2)}{2}\lambda_{1}^{2}+(2N^{2}+3N-6)\lambda_{1}\lambda_{2}+\frac{(N^{4}+5N^{3}-6N^{2}-36N+72)}{8}\lambda_{2}^{2}\right]\frac{15N_{f}N}{2048}\alpha_{g}\notag\\
&-\frac{(N^{2}+N+2)(N^{2}+N+14)}{2048}\lambda_{1}^{3}-\frac{3(N^{2}+N+14)(2N^{2}+3N-6)}{2048}\lambda_{1}^{2}\lambda_{2}\notag\\
&-\frac{3(7N^{4}+27N^{3}-33N^{2}-144N+252)}{2048}\lambda_{1}\lambda_{2}^{2}
-\frac{(2N^{6}+19N^{5}+21N^{4}-270N^{3}+108N^{2}+2160N-3024)}{8192}\lambda_{2}^{3},\\
\Delta_{\nu^{-1}}^{(3)}={}
&-\frac{N_{f}^{2}N^{2}(64 N^2+89 N-178)}{8192}\alpha_{g}^{3}\notag\\
&+\frac{N_{f}N(187 N^4 + (934 - 240 \zeta_{3}) N^3 -(912 \zeta_{3}-31) N^2 - 
 12 (377-144 \zeta_{3}) N+(4512-1728\zeta_{3}))}{32768}\alpha_{g}^{3}\notag\\
&+\left[(N^{2}+N+2)\lambda_{1}+(2N^{2}+3N-6)\lambda_{2}\right]\frac{3N_{f}^{2}N^{2}}{2048}\alpha_{g}^{2}\notag\\
&-\frac{N_{f}N(41 N^4 + 12 (2 \zeta_{3} + 7) N^3 + 9 (40 \zeta_{3} - 1) N^2 + 
 72 (8 \zeta_{3} - 1) N-4 (312 \zeta_{3}+7))}{8192}\lambda_{1}\alpha_{g}^{2}\notag\\
&-\frac{N_{f}N((48 \zeta_{3}+76) N^4+(288 \zeta_{3}+179) N^3-(312 \zeta_{3}+253) N^2-12 (168 \zeta_{3}+25) N+(3744 \zeta_{3}+84))}{8192}\lambda_{2}\alpha_{g}^{2}\notag\\
&-\left[\frac{(N^{2}+N+2)}{2}\lambda_{1}^{2}+(2N^{2}+3N-6)\lambda_{1}\lambda_{2}+\frac{(N^{4}+5N^{3}-6N^{2}-36N+72)}{8}\lambda_{2}^{2}\right]\frac{33N_{f}N}{2048}\alpha_{g}\notag\\
&-\frac{3(N^{2}+N+2)(5N^{2}+5N+64)}{1024}\lambda_{1}^{3}
-\frac{9(2N^{2}+3N-6)(5N^{2}+5N+64)}{1024}\lambda_{1}^{2}\lambda_{2}\notag\\
&-\frac{3(N^{6}+6N^{5}+789N^{4}+2980N^{3}-3660N^{2}-15840N+27648)}{8192}\lambda_{1}\lambda_{2}^{2}\notag\\
&-\frac{3(10N^{6}+92N^{5}+93N^{4}-1275N^{3}+522N^{2}+9936N-131824)}{4096}\lambda_{2}^{3}.
\end{align}
Note that the actual universal quantities are obtained by substituting the fixed point values into the expressions.

The three-loop order corrections in the universal quantities for $N=8$, as a representative example, are fitted by
\begin{align}
    \Delta_{\eta_{\psi}}^{(3)}(N=8,x)\approx{}&-\frac{0.814 x^2+142.170 x+1023.120}{x^3+75.176 x^2+1828.582 x+20034.194}\epsilon^{3}+\mathcal{O}(\epsilon^{4}),\\
    \Delta_{\eta_{\varphi}}^{(3)}(N=8,x)\approx{}&\frac{2.073 x^2+764.158 x+28101.110}{x^3+450.301 x^2+11033.469 x+225786.313}\epsilon^{3}+\mathcal{O}(\epsilon^{4}),\\
    \Delta_{\nu^{-1}}^{(3)}(N=8,x)\approx{}&-\frac{1.643 x^4+39.592 x^3+201958.980 x^2+5.236\times 10^6 x+2.654\times 10^7}{x^5+384.493 x^4+38049.982 x^3+1.116\times 10^6 x^2+1.531\times 10^7 x+4.986\times 10^7}\epsilon^{3}+\mathcal{O}(\epsilon^{4}),\\
    \Delta_{\mathcal{R}_{G}}^{(3)}(N=8,x)\approx{}&-\frac{103.338 x^3+25616.692 x^2+799467.484 x}{x^4+570.375 x^3+26532.963 x^2+799833.432 x+3.532\times 10^6}\epsilon^{3}+\mathcal{O}(\epsilon^{4}).
\end{align}
Note that these functional forms works well even for other $N$ and $x$.
The behaviors of the three-loop order corrections in the universal quantities for $x\gg N\gg1$ are given by
\begin{align}
    \Delta_{\eta_{\psi}}^{(3)}(N,x)\approx{}&-\frac{0.093 N x^2 + 1.730 N^2 x + 1.422 N^3}{x^3 + 7.186 N x^2 + 16.287 N^2 x + 17.338 N^3}\epsilon^{3}+\mathcal{O}(\epsilon^{4}),\\
    \Delta_{\eta_{\varphi}}^{(3)}(N,x)\approx{}&\frac{0.186 N x^2 + 1.928 N^2 x + 15.658 N^3}{x^3 + 12.198 N x^2 + 41.001 N^2 x + 58.950 N^3}\epsilon^{3}+\mathcal{O}(\epsilon^{4}),\\
    \Delta_{\nu^{-1}}^{(3)}(N,x)\approx{}&-\frac{0.188 N x^4 -2.811 N^2 x^3 + 220.088 N^3 x^2 + 714.782 N^4 x + 424.651 N^5}{x^5 + 405.821 N x^4 + 405.821 N^2 x^3 + 1220.237 N^3 x^2 + 1670.150 N^4 x + 613.029 N^5}\epsilon^{3}+\mathcal{O}(\epsilon^{4}),\\
    \Delta_{\mathcal{R}_{G}}^{(3)}(N,x)\approx{}&-\frac{11.905 N x^3 + 209.135 N^2 x^2 + 1217.245 N^3 x}{x^4 + 55.754 N x^3 + 297.681 N^2 x^2 + 1152.125 N^3 x + 653.140 N^4}\epsilon^{3}+\mathcal{O}(\epsilon^{4}).
\end{align}

\end{widetext}

\section{Borel-Pad\'{e} approximation}\label{app:borel-pade}
In this section, we will introduce the Borel-Pad\'{e} approximation.
The Borel-Pad\'{e} approximation is a combination of the Borel summation and Pad\'{e} approximation \cite{LeGuillou1977,goettkerschnetmann1999,Zinn-Justin}. The procedure is as follow:
As an example, let us consider the following expansion in terms of $\epsilon$ up to $\epsilon^{2}$ order.
\begin{align}
f(\epsilon)={}&a_{0}+a_{1}\epsilon+a_{2}\epsilon^{2}+\mathcal{O}(\epsilon^{3}).
\end{align}
First, the Borel transformation of $f(\epsilon)$ is given by
\begin{align}
\mathcal{B}[f](\epsilon)={}&\sum_{n=0}^{\infty}\frac{a_{n}}{n!}\epsilon^{n}\notag\\
={}&a_{0}+a_{1}\epsilon+\frac{a_{2}}{2}\epsilon^{2}+\mathcal{O}(\epsilon^{3}).\label{eq:borel_tr}
\end{align}
Then, we apply the Pad\'{e} approximation into Eq.~\eqref{eq:borel_tr}. In particular, we will consider the Pad\'{e} approximation of order $[1/1]$, here. It is given by
\begin{align}
    \mathcal{P}_{[1/1]}[\mathcal{B}[f](\epsilon)]={}&\frac{\bar{a}_{0}+\bar{a}_{1}\epsilon}{1+\bar{b}_{1}\epsilon},
\end{align}
where
\begin{align}
\bar{a}_{0}={}a_{0},\quad
\bar{a}_{1}=a_{1}-\frac{a_{0}a_{2}}{2a_{1}},\quad
\bar{b}_{1}={}-\frac{a_{2}}{2a_{1}}.
\end{align}
Finally, we get the Borel-Pad\'{e} approximation for a given expression by performing the Borel integral transformation, 
\begin{align}
\tilde{f}(\epsilon)={}&\int_{0}^{\infty}e^{-t} \mathcal{P}_{[1/1]}[\mathcal{B}[f](t\epsilon)]dt.
\end{align}

%

\end{document}